\documentclass[pre,twocolumn,groupedaddress]{revtex4-1}
\usepackage{graphicx} 

\begin{document}


\title{Opposed flow focusing: evidence of a second order jetting transition}


\author{Jun Dong,\textit{$^{a,b}$} {Max Meissner,\textit{$^{a,b}$} Jens Eggers,\textit{$^{c}$}} Annela M. Seddon,\textit{$^{a,b,d}$} and C. Patrick Royall \textit{$^{a,b,e}$}}
\email[Electronic address: ]{paddy.royall@bristol.ac.uk}
\affiliation{$^{a}$~H.H. Wills Physics Laboratory, University of Bristol, Bristol, BS8 1TL, UK}
\affiliation{$^{b}$~Centre for Nanoscience and Quantum Information, Tyndall Avenue, Bristol, BS8 1FD, UK}
\affiliation{$^{c}$~Mathematics Department,University of Bristol, BS8 1TW, Bristol UK}
\affiliation{$^{d}$~Bristol Centre for Functional Nanomaterials, University of Bristol, Bristol, BS8 1TL, UK}
\affiliation{$^{e}$~Chemistry Department, University of Bristol, Bristol, BS8 1TS, UK}


\date{\today}

\begin{abstract}
We propose a novel microfluidic ``opposed-flow'' geometry in which the continuous fluid phase is fed into a junction in a direction opposite the dispersed phase. This pulls out the dispersed phase into a micron-sized jet, which decays into micron-sized droplets. As the driving pressure is tuned to a critical value, the jet radius vanishes as a power law down to sizes below 1 $\mu$m. By contrast, the conventional ``coflowing'' junction leads to a first order jetting transition, in which the jet disappears at a finite radius of several $\mu$m, to give way to a ``dripping'' state, resulting in much larger droplets. We demonstrate the effectiveness of our method by producing the first microfluidic silicone oil emulsions with a sub micron particle radius, and utilize these droplets to produce colloidal clusters.
\end{abstract}


\maketitle

\section{Introduction}
The controllable production of  micron scale emulsions is an area of
highly active research. Allowing extreme reductions in sample volumes,
microdroplets currently find use across both fundamental research and
industrial applications. One promising manufacturing method for these
droplets is microfluidic emulsification~\cite{theberge2010}. Offering
unparalleled control over droplet formation, microfluidic emulsification
systems find use across applications as diverse as the generation of
artificial cells~\cite{martino2016}, high-throughput screening of
patient samples~\cite{beebe2002, guo2012, yi2006, delamarche2005},
colloidal model systems~\cite{meissner2017}, or even traffic
dynamics~\cite{champagne2010traffic}.

\begin{figure}[h!]
\centering
\includegraphics[width=\linewidth]{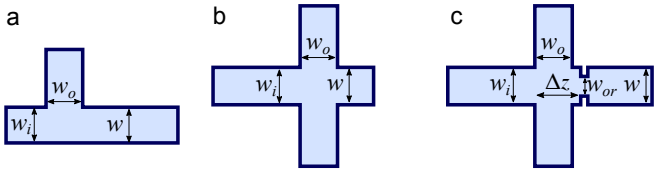}
\caption[device types]{Schematics of: (a) T-junction device, (b) a coflow device, and (c) a flow focusing device.}
\label{DType}
\end{figure}

Conventionally, droplet generation in a microfluidic device is
achieved via three main geometries as shown in Fig.~\ref{DType}:
(a) T-junctions in which viscous shear stress at one fluid interface
pulls off droplets into the flow of a second immiscible
fluid~\cite{thorsen2001}, (b) coflowing devices where an outer continuous
phase fluid flows parallel to and surrounding an inner dispersed phase
fluid until droplet generation occurs via stretching of the fluid
interface~\cite{ganan-calvo2001}, and flow-focusing devices, where
the interface between coflowing streams are forced through a flow
constriction causing droplet breakup through the generation of a
velocity gradient~\cite{anna2003}.

However, while these methods can effectively produce droplets with
sizes below 100 $\mu$m, the production of true micro-droplets with
sizes below 10 $\mu$m in diameter via microfluidic means remains a
significant challenge. Typical droplet sizes realized in previous
microfluidic setups are summarized in Table~\ref{size_table}. 
To produce smaller droplets, one possibility
is to reduce the size of the microfluidic device, but this is limited
by the resolution of the manufacturing process available~\cite{garstecki2005}.
In addition, small channel sizes lead to large velocity gradients and
large pressure gradients needed to drive the flow. In particular,
oil-in-water emulsification remains a largely unexplored field. 
	
\begin{table*}[t]
\centering
\small
\caption{Comparison of microfluidic droplet generation methods}
\label{tbl:example}
\begin{tabular*}{\textwidth}{@{\extracolsep{\fill}}llll}
\hline
Droplet generation method & Emulsion type & Water droplet diameter &
Oil droplet diameter\\
  \hline
 T-junction droplet generation & Water-in-oil~\cite{nisisako2002} or
 oil-in-water~\cite{nisisako2005} & 50 $\mu$m & 100 $\mu$m\\
 Capillary coflow droplet generation & Water-in-oil or
 oil-in-water~\cite{haase2014} & 20 $\mu$m & 10 $\mu$m\\
 Flow focusing & Water-in-oil~\cite{anna2003} or
 oil-in-water~\cite{bauer2011} & 10 $\mu$m & 20 $\mu$m\\
 Partial wetting & Fluorinated oil in water~\cite{cohen2014} & -& 6 $\mu$m \\
Geometric break-up & Water in oil~\cite{link2004} & 50 $\mu$m & -\\
Tip-streaming & Water in oil~\cite{jeong2012} & 3 $\mu$m & -\\
\hline
\end{tabular*}
\label{size_table}
\end{table*}

Attempts to overcome this size limitation include methods of switching
to more robust polymeric materials from which the device
is comprised such as Norland adhesives to allow smaller
geometries~\cite{meissner2017}, using multiple coaxial jets
\cite{ganan-calvo2007}, or
utilizing electric fields to induce electro-hydrodynamic
jetting~\cite{G97,HLGVMP12}. While each of these methods begins to
address the problem in their individual way, they all carry significant
limitations in terms of technical complexity or system inflexibility.
Here we pursue the alternative route of using suitably chosen flow
characteristics to focus the inner phase into a very thin jet, whose
radius is no longer limited by the device size. 

Droplet production from a highly focused jet, also known as tip streaming, 
is a flow mode in which a thin jet emerges from a nearly conical point
\cite{taylor1934,suryo2006,castrohernandez2012,anna2016}. The jet is
subject to the Rayleigh-Plateau instability \cite{EV08}, and decays into
droplets further downstream, whose sizes are set by the radius of the jet.
However, the circumstances under which such a jet is produced are not
understood, and depend crucially on the confined flow conditions
realizable in a microfluidic device \cite{anna2016}. 
In unbounded flows produced by a
four-roll mill, G.I. Taylor\cite{taylor1934} was able to deform the end
of a drop into a conical tip, yet tip streaming occurred only when a
small amount of surfactant was added. As the system was drained of surfactant,
the jet disappeared once again. 

A similar phenomenon was observed in the selective withdrawal geometry,
in which an upper fluid phase is withdrawn through a nozzle
from near the interface between two fluids layered atop of one another
\cite{cohen2002,case2007}. As the flow rate is slowly increased, the
interface is deformed into an increasingly sharp ``hump''. When the
tip of the hump has reached a radius of curvature of about 200 $\mu$m,
it transitions toward a ``spout'', in which a thin jet is entrained
into the nozzle; with increasing flow rate, the jet becomes thicker.

However, there is hysteresis in the system: one needs to go to lower flow rates
for the spout to disappear than those at which the spout was first formed. 
Thus this system bears the characteristics of a first order transition:
firstly, it is discontinuous, in that the jet radius jumps from zero
(no jet) to a finite value, and vice versa. Secondly, there is hysteresis,
in that the values of the control parameter are different depending on
whether one passes from jet to no jet or vice versa.
Both the hump and the spout states are characterized by a power-law
dependence of their characteristic sizes as the control parameter
approaches a critical value. This indicates that one is close to a
second order transition, in which the characteristic size goes to
zero, so that the transition from jet to no jet is continuous.

The aim of this paper is to use the precise control over the flow field
made possible by microfluidics in order to realize this hypothetical
second order transition. As the difference between the two states
vanishes at a second order transition, we also expect there to be no
hysteresis in that case. Achieving a second order transition would
mean that the length scale of the jet is no longer set by the
size of the microfluidic setup, but rather by our ability to tune the flow parameters close
to the transition. 

Recently, there has been some progress describing the formation of 
narrow tips and thin jets in microfluidic devices, both experimentally
and computationally \cite{anna2016}. Usually this is achieved by extracting
the dispersed fluid from a nozzle with an exterior phase flowing in the
same direction (the coflowing configuration). In this situation the
``jetting'' state, which allows for the formation of the smallest drops,
competes with a ``dripping'' state, characterized by the periodic
formation of individual drops from near the nozzle opening \cite{utada2007}.

This is what is known as an ``absolute instability''
\cite{Cho05,UFGW08}, in that breakup occurs in the frame of reference 
of the nozzle. By contrast, drop formation from the jet occurs by a
``convective instability'', which grows in a frame of reference convected
with the flow. Numerical calculations in simple flow geometries have
confirmed the possibility of creating thin jets from the tip of conical
points \cite{suryo2006,castrohernandez2012}. However, jet radii have
not been reported in a systematic fashion; in particular, the crucial 
question of what limits the size of the smallest jet has not been addressed.
	
In this paper, we contrast jetting in the conventional coflowing geometry
with a novel opposed flow geometry. In agreement with previous results,
we find in the former case that the smallest jet size is limited by transition
to a dripping state. In the opposed flow geometry, on the other hand,
dripping is suppressed, and the smallest jet size of about 1$\mu$m
appears to be limited only by our ability to adjust parameters close to a
second order transition. This represents a significant improvement over
droplet sizes of 10 $\mu$m typically available from conventional flow
focusing and opens the way to even smaller droplets than those we have produced here. 
The droplets produced this way are small enough so that they are
susceptible to the thermal energy of the system, i.e. they are colloidal.
Assemblies of such droplets have a well-defined thermodynamic state and
thus can reach their ground state. 
	
One experimental system where the size of these droplets can be
readily exploited is the study of colloidal clusters. Displaying
structural ordering rather different from that of bulk materials,
colloidal clusters represent one of the clearest links between local
geometry and bulk condensed matter~\cite{malins2013tcc}. In particular,
due to the five-fold symmetry found in structures such as icosahedral and
decahedral, colloidal clusters can act as model systems for both
biological systems such as viral capsids, or materials such as glass
where colloidal model systems have led to great insight into how atoms
self-organize into energy minimizing locally favored
structures~\cite{malins2009,meng2010,malins2013tcc,klix2013}. We demonstrate the utility of the droplets
produced via our opposed  flow-focusing methods by producing colloidal
clusters of droplets via the addition of a depletion potential.

This paper is organized as follows: in Sect.~\ref{MaterialsMethods}
the experimental details and protocols will be described. In
Section~\ref{ResultsDiscussion} we contrast the two flow geometries,
and discuss their scaling properties in detail. As a potential application,
we demonstrate the formation of colloidal clusters. We conclude by
summarizing our findings and exploring their implications. 
	
\section{Materials and Methods} \label{MaterialsMethods}
\subsection{Device assembly}
Device patterns were fabricated on silicon wafers using standard
photolithographic methods. Etched wafers were treated with
trichloro(1H,1H,2H,2H-perfluorooctyl) silane to allow easy siloxane lift-off.
Polydimethylsiloxane(PDMS) was mixed up in a 10:1 ratio of elastomer to
curing agent (Sylgard 184). The mixed silicone elastomer was degassed,
and approximately 20 g was poured on the silicon wafers, degassed a
second time to remove any remaining bubbles, and heat cured at 60~$^{\circ}$C
for 6 hours. Once cured, PDMS layers were cut to shape and carefully
removed from the silicon wafer. Tubing connectors were punched into the
PDMS slabs using a $0.8$~mm diameter biopsy punch. The cut and punctured
PDMS was subsequently thoroughly washed with isopropanol to clean out any
PDMS remnants from the puncturing process. The PDMS chips were then plasma
treated for 30 s in a 100 W Diener plasma cleaner. Immediately following
plasma treatment, the device chip was brought into contact with a similarly
prepared glass  microscope slide, bonding the activated surfaces together.

We consider two different flow geometries, as shown in Fig.~\ref{AvO},
produced from the same device with a central straight oil channel,
to which four water side channels are attached in an x-shaped configuration.
Two of the side channels are not punctured during the fabrication process,
and therefore are blocked when liquid flows through the device. In the
opposed flow system, shown in Fig.~\ref{AvO}(a), the oil channel is
connected so that oil flows from the side of the blockage, so that
the water channel makes a 165$^{\circ}$ angle with the oil channel. 
Figure~\ref{AvO}(b) shows the coflowing system, in which the oil comes
from the direction of the open channels, leading to an angle of
15$^{\circ}$ between the aqueous and the oil flow.

\subsection{PDMS surface coating}
The hydrophobic surface properties of PDMS lead to clogging when
producing oil-in-water emulsions. We overcome this difficulty by 
covering the PDMS with several layers of polymer coating~\cite{bauer2010},
which were applied by flowing alternating polymer electrolyte solutions
through the channel by use of a syringe pump. Layers were applied as follows:
first 4 $\mu$l of poly(allylamine hydrochloride) (PAH), a positively
charged electrolyte, was inserted into the channel, so that it
coated the negatively charged PDMS surfaces. This was followed by
depositing the same volume of poly(sodium 4-styrenesulfonate) (PSS),
a negatively charged electrolyte, onto the PAH layer. In total, four
layers of PAH and PSS were applied to the PDMS surfaces. In between
polymer electrolyte segments, 2 $\mu$l of NaCl solution was flushed
through the channels to remove excess charge. The solutions of PAH and
PSS were used at the same concentration of 0.1 \% w/v in 0.5 M NaCl, and
the concentration of the NaCl solution used for washing in between the
polymer segments was 0.1 M. These segments were loaded into a length
of tubing, and flushed through the PDMS device at a flow rate of 50 $\mu$l/h.

\subsection{Droplet and jet production}
Jetting experiments were carried out by connecting the assembled PDMS
chip to a pressure pump (Fluigent MFCS) with polyethylene tubing.
Silicone oil (shear viscosity 4.57 mPa$\cdot$s) dyed with nile red was
used as the dispersed phase and the continuous phase was a mixture of
water and glycerol (30 : 70 wt$\%$) with viscosity 23 mPa$\cdot$ s. This
leads to a viscosity ratio of outer to inner phase $\eta_o / \eta_i = $ 5.0. 
In order to produce a stable jet, we reduced the surface tension by adding
21mM of sodium dodecyl sulfate (SDS) to the aqueous phase, which is the
maximum concentration that would go in solution. Since this is several
times the critical micelle concentration (CMC) \cite{CDA08}, we believe 
that replacement of surfactant at the interface is so fast that the surface
tension can be considered constant \cite{anna2016}. 

\begin{figure}[h!]
\centering
\includegraphics[width=\linewidth]{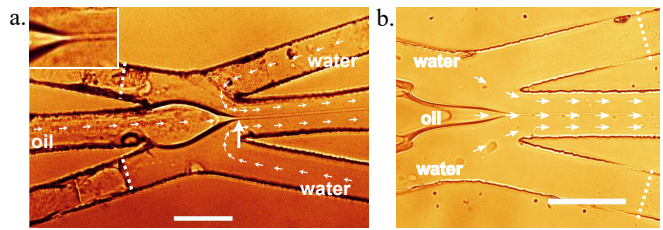}
\caption[Comparison of opposed and coflow flow regimes]
{Schematics of: (a) opposed flow system with an angle of 165 $^{\circ}$
between the aqueous and oil flow. Inset shows close-up of jet.
(b) coflowing system with an angle of 15 $^{\circ}$ between the aqueous
and the oil flow. Dashed lines show blocked channels. Scale bars
represent 75 $\mu$m.}
\label{AvO}
\end{figure}

\begin{figure}[htb!]
\centering
\includegraphics[width=1\linewidth]{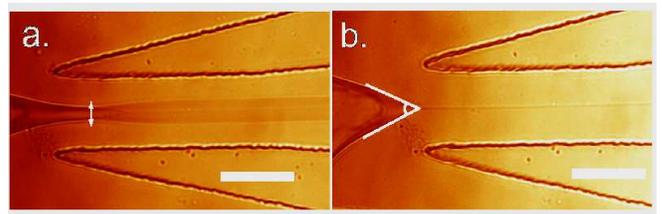}
\caption[Measurement points for conical jets] {Images
of microfluidic devices with opposed flows showing measurement method
for (a) the jet width and (b) the cone angle. The pressures of the
liquid phases are 700 mbar external, 350 mbar internal, and 700 mbar
external, 309 mbar internal, (a) and (b) respectively.
Scale bars represent 50 $\mu$m.}
\label{Quadrant}
\end{figure}

\begin{figure*}[htb!]
\centering
\includegraphics[width=0.8\linewidth]{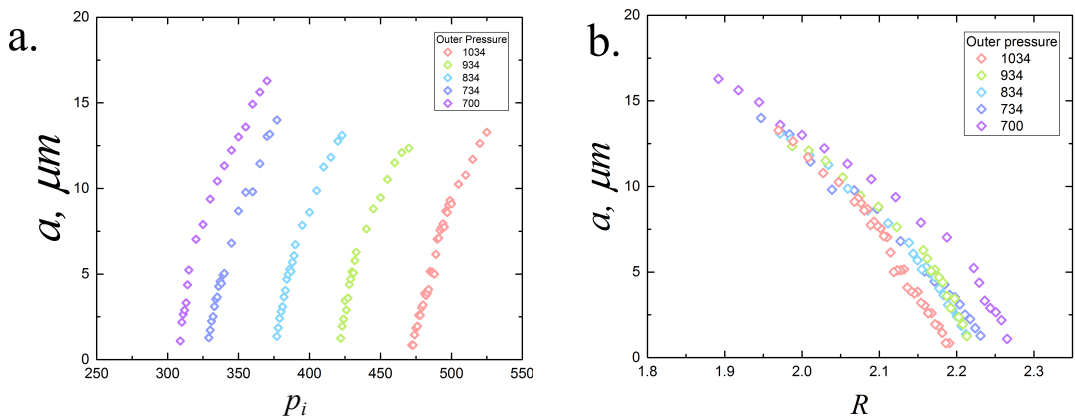}
\caption[Pressure vs pressure ratio]{Jet radius $a$ as function of the outer
and inner pressures in the opposed flow system. All pressures are
given in mbar (a) jet radius versus inner pressure for various
outer pressures. (b) same data plotted against the pressure ratio
$R = p_i/p_o$.}
\label{Qupvr}
\end{figure*}

The driving pressure was controlled using MaesFlo 3.2 software
(Fluigent, Paris, France). Pressures were adjusted until the oil flow
was stable and producing droplets, as shown in Fig.~\ref{AvO} for
the opposed flow (left) and coflowing geometries (right). Once stable
droplet formation was achieved, flow parameters were adjusted by either
holding the external pressure constant and varying the internal pressure,
or vice versa. The former mode of control is illustrated in
Figs.~\ref{Quadrant} and \ref{Qupvr} for the opposed flow geometry.
As the internal pressure is lowered, the jet radius becomes very thin,
and the oil drop near the nozzle exit assumes an almost conical shape,
from which the jet emerges. 
In Fig.~\ref{Qupvr}(a), each color represents the jet radius $a$
as the inner phase pressure is decreased, while the outer phase pressure
was held fixed at a certain value from 1034 to 700 mbar. In
Fig.~\ref{Qupvr}(b) we show that all curves can be brought to a near
collapse by plotting the radius as a function of the pressure ratio
$R = p_i/p_o$, establishing that this ratio is the main parameter
determining the state of the system. 

\subsection{Cluster formation}
Droplets obtained via the opposed flow focusing geometry were formed into
clusters via the addition of non-adsorbing polymers. To this end
droplets approximately 1.5 $\mu$m 
in radius were produced, with the
aqueous phase containing 6.34 mg/ml hydroxyethyl cellulose (HEC). With
the radius of gyration of HEC being 50 nm, this leads to a colloid to
polymer size ratio of approximately $q\approx0.03$. Collection of the droplets
was carried out with a fixed inner phase pressure of 275 mbar with the
aqueous flows kept at 640 mbar. The resulting  droplets were collected
into a glass capillary and imaged under a confocal microscope
(Leica SP-8
) with excitation wavelength of 514 nm.

\section{Results and discussion}
\label{ResultsDiscussion}
Below we contrast the two geometries of coflow and opposed flow. 
In the coflowing geometry, the smallest attainable jet gives way 
to dripping, leading to a hysteretic first order transition. In the
opposed flow geometry, the smallest jet size is limited only by our
ability to control the pressure, leading to a state where flow ceases.
Oil drops with radii below 1 $\mu$m, can be produced by this process. 
Here we show that such colloidal-sized droplets
assemble into 
clusters with the addition of non-adsorbing polymers. 

\subsection{Coflowing jet formation}
We begin with the conventional coflowing orientation in order to
produce an oil jet from the drop attached to the nozzle, the
aqueous flow channels making a 15$^{\circ}$ angle with the oil channel. 
By controlling the pressure ratio $R=p_i/p_o$ as
detailed in the method section above, stable droplet formation was
induced in the microfluidic device. To characterize the size scaling
with pressure ratio, the radius at the narrowest point of the jet was measured. 

\begin{figure*}[htp!]
\centering
\includegraphics[width=0.8\linewidth]{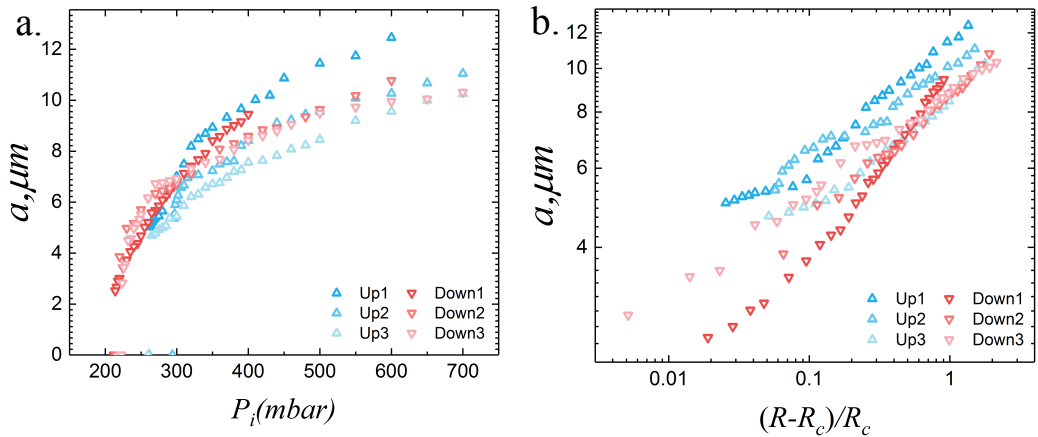}
\caption[Figure showing behavior of normal flow focusing system]
{In the coflowing geometry, the outer phase pressure was fixed at
$p_o = 700$ mbar, the jet radius was measured while varying the inner
oil phase pressure. 'Up' refers to an experimental run during which
the inner pressure increases (increasing the jet radius), 'down'
where the inner pressure decreases (decreasing the jet radius).
'Up' and 'down' processes were repeated three times to check the stability
of the data. (a) jet radius as a function of the inner oil pressure
$p_i $, (b) log-log plot of the jet radius as function of 
$(R-R_c)/R_c$. }
\label{NormFF}
\end{figure*}

The raw data of jet radius $a$ as a function of the inner flow (oil)
pressure $p_i$ is shown on the left; the outer, aqueous pressure
was fixed at 700 mbar. Shown are three cycles during which $p_i$ increases
from zero until a jet is formed and subsequently increases in radius
(blue symbols), followed by a sequence of measurements for decreasing
$p_i$ (red symbols). There is considerable scatter in the data between
each cycle. In addition, for each cycle there is significant hysteresis in
that a jet forms at a radius of about 5$\mu$m, while the jet disappears
only when the radius has decreases to about 2$\mu$m. Just below the transition,
the system is in the dripping mode, in which oil drops are produced
directly from the nozzle in a periodic fashion. Upon further decrease
of $p_i$, flow stops completely and the interface between oil and the
aqueous phase assumes a rounded shape. 

To test whether there is some indication of scaling in the coflowing
data, in Fig.~\ref{NormFF}(b) we plotted the same radius data in a
log-log plot as function of $(R-R_c)/R_c$. For each
cycle, and for each data set going up or down, we adjusted a critical
pressure ratio $R_c$ such that we obtained an optimal power-law fit
\begin{equation}
a = A (R - R_c)^{\alpha},
\label{pl}
\end{equation}
where $A$ and the exponent $\alpha$ were also adjustable parameters. 
This means that $R_c$ is the value of the pressure ratio at which the
jet radius would vanish in a second order (continuous) transition.
However, similar to measurements in the selective withdrawal geometry
\cite{cohen2002,case2007}, scaling is cut off at finite $a$ in a discontinuous
transition. Scaling exponents also give inconsistent values and were found
in the ranges $\alpha=0.2-0.24$ in the ``up'' direction, and
$\alpha=0.22-0.37$ in the ``down'' direction. 

\subsection{Opposed flow jet formation}

\begin{figure*}[htp!]
\centering
\includegraphics[width=0.8\linewidth]{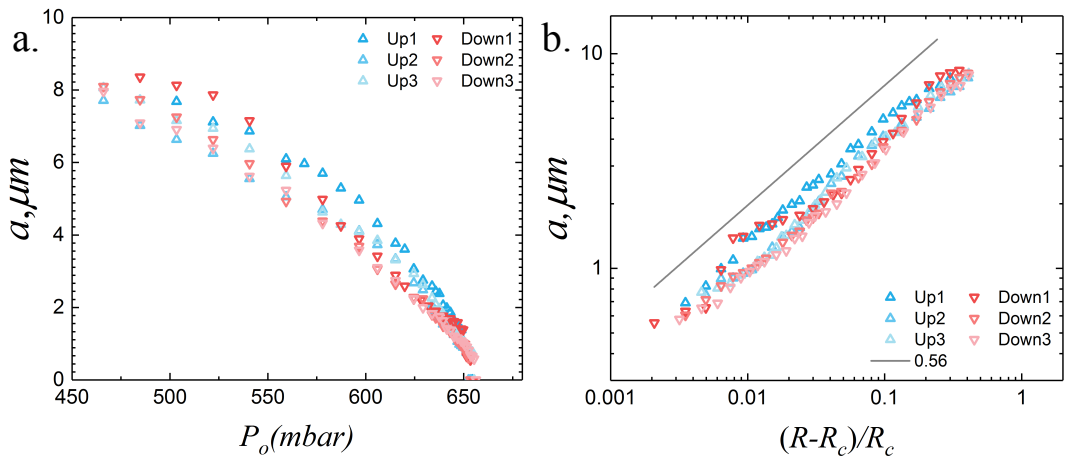}
\caption[Figure showing behavior of opposed flow focusing system]
{In the opposed flow geometry, the inner oil phase pressure was fixed
at $p_i=275 $mbar, the jet radius was measured while varying the outer
phase pressure. 'Up' refers to an experimental run during which
the outer pressure decreases (increasing the jet radius), 'down'
where the outer pressure increases (decreasing the jet radius).
(a) jet radius as a function of the outer aqueous pressure $p_o$,
(b) log-log plot of jet radius as a function of $(R-R_c)/R_c$.
The line in (b) has a slope of 0.56.}
\label{RevFF}
\end{figure*}

By inverting the flow direction of the oil, we now report results for the
novel opposed flow geometry, in which the angle between the aqueous
flow channel and the oil flow channel is 165$^{\circ}$.
In Fig.~\ref{RevFF}(a) we show the jet
radius for three cycles of increasing jet radius and decreasing jet
radius. This time, the inner oil phase pressure was fixed at 
$p_i = 275$ mbar, while the outer pressure $p_o$ was varied. Hence
during the ``up'' phase, $p_o$ is decreased so as to increase $R$,
while during the ``down'' phase $p_o$ is increased. In the opposed flow
geometry, there is very little hysteresis, and there is a much better
collapse of the data across the three cycles of varying the outer pressure.
This is even clearer in the log-log plot of Fig.~\ref{RevFF}(b), 
where for each cycle and for each direction of increasing and decreasing
$p_o$, we fitted the data to the power law (\ref{pl}). With the
critical value $R_c$ in hand, we plotted $a$ against the critical
pressure parameter $(R-R_c)/R_c$.

There is now little variation of the slope in each case; in the
``up'' direction we obtain values $\alpha = 0.55-0.58$, in the
``down'' direction $\alpha = 0.54-0.57$. Fitting to the data
for all three cycles leads to an average exponent of $\alpha = 0.56$,
shown as the straight line in Fig.~\ref{RevFF}(b). The smallest
jets produced in the opposed flow geometry have a radius of about
0.5$\mu$m, significantly smaller than anything produced by conventional
focusing, see Table~\ref{size_table}. One example of collected droplets from the device was characterized under the confocal microscope and is shown in Fig. \ref{Clusters}(a), with mean radius of 1.2$\mu$m. We were unable to determine
whether there remained a small discontinuity as the jet disappears,
or whether our ability to produce a small jet is limited by the accuracy
with which the outer pressure can be adjusted. Most importantly,
in the opposed flow geometry we no longer see a dripping state, but
we pass directly from a jetting state to a state of no flow. 

\subsection{Droplet clusters}
\begin{figure}[h!]
\centering
\includegraphics[width=\linewidth]{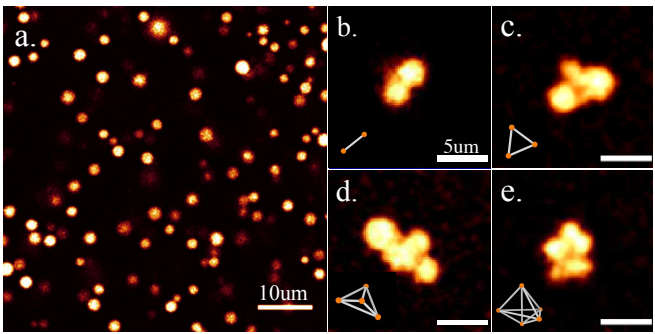}
\caption[Confocal images of collected droplets]{Confocal microscopy
images of collected droplets and clusters from the opposed flow
focusing device. (a) Colloidal emulsion droplets with mean radius of
1.2 $\mu$m, the scale bar represents 10 $\mu$m. (b) - (e) Collected
droplet clusters that are formed by depletion attraction of
non-absorbing polymers HEC. Clusters contain droplets from N=2 to N=5,
with wireframes indicating the geometry of clusters; scale bars
represent 5 $\mu$m.}
\label{Clusters}
\end{figure}

Since the jet decays into droplets further downstream, the very thin 
jets produced in the opposed flow geometry allow us to make correspondingly
smaller droplets, with hydroxy ethyl cellulose 
polymers added to them. Owing to the short-range
depletion attractions induced by HEC polymers, colloidal droplets with
mean radius of 1.5 $\mu$m form clusters, as shown in the confocal images
of Fig.~\ref{Clusters}(b)-(e). These enable studies in the same spirit as
\cite{meng2010}, but with 
3d imaging, as the system is refractive
index matched. In Figs.~\ref{Clusters}(b)-(e) we show closeups of individual
clusters with $N=2-5$ particles; the wireframe inserts indicate the
geometry of the cluster.

Our new model system enables the study of near-frictionless droplets.
Unlike systems with solid particles, where access to the thermodynamic
ground state is often suppressed~\cite{klix2013,royall2018thunderbox},
perhaps due to the polymer stabilizer layer thickness being the same
size as the depletion polymers ~\cite{prasad2003}, these near-frictionless
droplets may be better able to reach the ground state. The potential
to generate smaller droplets with this method further enables us to
form clusters with less tendency to sediment, and which may explore
their configuration space very quickly, as the colloid diffusion time
scales with the cube of the diameter. With a higher volume fraction
of droplets, percolating colloidal gels can be obtained through the
same depletion mechanism \cite{zaccarelli2007,manley2005,griffiths2017}.

\section{Conclusions}
\label{Conclusion}
In conclusion, it remains difficult to make sub 10 $\mu$m droplets
using conventional microfluidic methods, particularly in the case of
oil-in-water systems. Instead, we have described a method to make
colloidal oil droplets by tuning close to a second order transition,
thereby reducing the droplet size by almost an order of magnitude.
This was achieved by simply reversing the direction of the oil flow.
If one succeeds in coming even closer to the transition, our approach
has the potential of reducing droplet size practically without limit. 
We demonstrate the utility of the droplets produced here via the
production of colloidal clusters, an experimental system which is
challenging to access otherwise.

\bibliographystyle{unsrt}

\begin{acknowledgments}
J.D. acknowledges Bayer CropScience AG for financial support. M.M. acknowledges the EPSRC Doctoral training allowance EP/L504919/1,
CPR acknowledges the Royal Society and Kyoto University SPIRITS fund, J.D. and CPR acknowledge the European Research Council (ERC consolidator grant NANOPRS, project number 617266) for financial support. J.E. acknowledges support
from the Leverhulme Trust through International Academic
Fellowship IAF-2017-010. He also gratefully acknowledges 
discussions with J.K. Nunes, as well as helpful comments on the manuscript. 
\end{acknowledgments}

\bibliography{GoWithTheFlow.bib}

\end{document}